%% file: patrick_xie_nuphys2016_proceedings.tex
\documentclass[12pt]{article}
\usepackage{graphicx}

\usepackage{subcaption}

\newcommand\pubnumber{arXiv:1704.06670}
\newcommand\pubdate{\today}

\def\ucl{Department of Physics and Astronomy\\
University College London, Gower St, London WC1E 6BT, UK}
\def\support{\footnote{On behalf of the SuperNEMO collaboration}}
\def\twonu{$2\nu\beta\beta$}
\def\zeronu{$0\nu\beta\beta$}
\def\bb{$\beta\beta$}

\def\Title#1{\begin{center} {\Large #1 } \end{center}}
\def\Author#1{\begin{center}{ \sc #1} \end{center}}
\def\Address#1{\begin{center}{ \it #1} \end{center}}

\newcommand\pubblock{\rightline{\begin{tabular}{l} \pubnumber\\
         \pubdate  \end{tabular}}}
\newenvironment{Abstract}{\begin{quotation}  }{\end{quotation}}
\newenvironment{Presented}{\begin{quotation} \begin{center} 
             PRESENTED AT\end{center}\bigskip 
      \begin{center}\begin{large}}{\end{large}\end{center} \end{quotation}}
\def\Acknowledgements{\bigskip  \bigskip \begin{center} \begin{large}
             \bf ACKNOWLEDGEMENTS \end{large}\end{center}}

\input econfmacros.tex

\begin{document}
\begin{titlepage}
\pubblock

\vfill
\Title{Status of the SuperNEMO 0$\nu\beta\beta$ experiment}
\vfill
\Author{ Cheryl Patrick and Fang Xie\support}
\Address{\ucl}
\vfill
\begin{Abstract}
SuperNEMO is an ultra-low-background
tracker-calorimeter experiment designed to look for the neutrinoless double-beta decay of 
various isotopes. We present the current state of the experiment's Demonstrator Module,
which is currently being installed and commissioned at the LSM in France.
\end{Abstract}
\vfill
\begin{Presented}
NuPhys2016, Prospects in Neutrino Physics\\
Barbican Centre, London, UK,  December 12--14, 2016
\end{Presented}
\vfill
\end{titlepage}
\def\thefootnote{\fnsymbol{footnote}}
\setcounter{footnote}{0}

\section{Introduction}

Double-beta decay  (\twonu{}), in which two neutrons decay simultaneously inside a nucleus, ejecting  two electrons and 2 electron antineutrinos, has been observed in several isotopes. However, if neutrinos are Majorana particles, it should be possible for a double-beta decay to occur with no neutrinos in the final state. These \zeronu{} events can be distinguished from \twonu{} decays by summing the energies of the two beta electrons: for \zeronu{}, the electrons will carry the full decay energy $Q_{\beta\beta}$, while for \twonu{} events, the summed electron energies will form a continuum up to  $Q_{\beta\beta}$. 

SuperNEMO \cite{supernemo} has been designed to study double-beta decays, and in particular, to look for \zeronu{} decays, which have never been observed. It builds on the design principles of its predecessor, NEMO-3 \cite{nemo3}, with an ultra-low-background tracker-calorimeter architecture, allowing us both to measure electron energies and to fully reconstruct particle tracks to identify \bb{} events. 

\begin{figure}[h]
\centering
\includegraphics[height=1.5in]{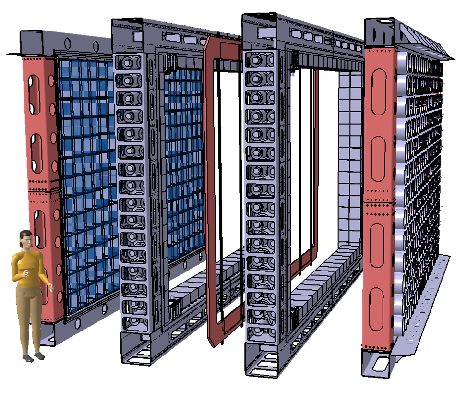}
\caption{Expanded diagram of the SuperNEMO Demonstrator showing (from left to right) calorimeter wall, tracker, source foil frame, tracker, calorimeter wall.
}
\label{fig:schematic}
\end{figure}

The SuperNEMO Demonstrator Module (figure \ref{fig:schematic}) has a layered design, with foils of \bb{} emitter sandwiched between tracker modules, surrounded by calorimeter walls. Initially, the source isotope will be 7kg of $^{82}$Se, mixed in a PVA base to create thin foils, suspended from the source foil frame. The modular design allows us to change these foils to study other isotopes.

The tracker, constructed in four C-shaped sections, consists of 2034 3-metre long drift cells operating in Geiger mode, arranged in rows of nine cells on each side of the source foil. Each cell comprises a central anode wire surrounded by 12 field-shaping wires, with copper cathode end caps at either end. When a charged particle crosses the cell, the anode signal timing tells us the distance from the anode wire, while the relative timings of the cathode signals give a position along the wire, allowing three-dimensional reconstruction.
 
The two calorimeter walls, situated outside the tracker, consist of 520 optical modules; 8-inch radiopure PMTs coupled to polystyrene scintillator blocks wrapped in teflon and mylar, with individual iron shielding. Lower-resolution optical modules around the edges of the tracker (giving a total of 712 modules) offer $4\pi$ acceptance. 

In its initial running period of 2.5 years, the Demonstrator Module will have a sensitivity to the \zeronu{}   half-life of $T_{1/2}^{0\nu\beta\beta}>6.5\times 10^{24}$ years, corresponding to a Majorana neutrino mass $\left< m_\nu \right> < 200-400$meV. A proposed 20-module full SuperNEMO detector with an exposure of 500 kg years (5 years, 100 kg of $^{82}$Se) would improve our sensitivity to $T_{1/2}^{0\nu\beta\beta}>10^{26}$ years ($\left< m_\nu \right> < 50 - 100$ meV).

\section{Tracker and calorimeter installation at LSM}

\begin{figure*}[h!]
    \centering
    \begin{subfigure}[t]{0.45\textwidth}
        \centering
        \includegraphics[height=1.5in]{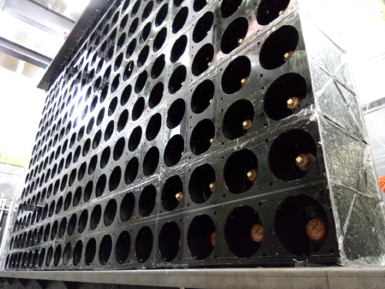}
        \caption{The back side of one of the main calorimeter walls}
        \label{fig:calo}
    \end{subfigure}
    \begin{subfigure}[t]{0.45\textwidth}
        \centering
        \includegraphics[height=1.5in]{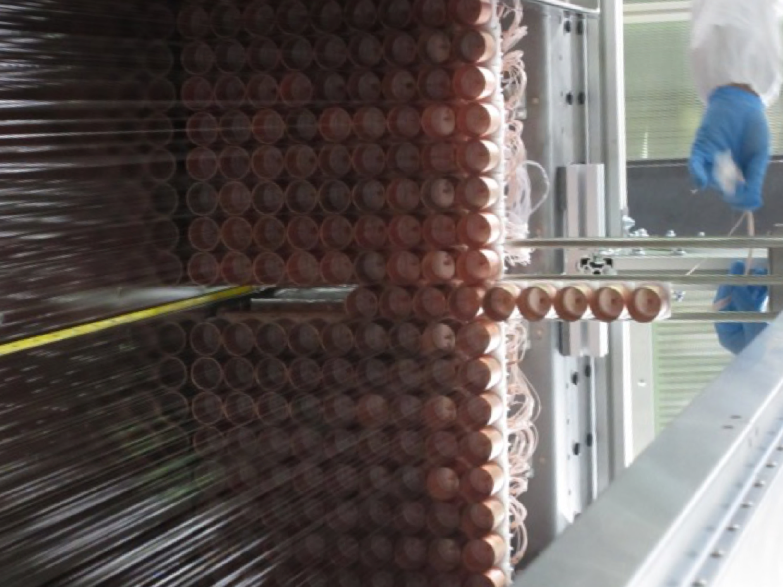}
        \caption{A final row of cells is inserted as two tracker C-sections are coupled}
        \label{fig:tracker}
    \end{subfigure}
    \caption{Installation of the Demonstrator Module at LSM}
\end{figure*}

The optical modules have all been built, and in 2016, the two outer calorimeter walls (figure \ref{fig:calo}) were assembled at the Laboratoire Souterrain de Modane (LSM), in the Fr\'{e}jus road tunnel in France. 
The four C-shaped tracker sections were constructed and commissioned in the UK, and by December 2016, all four had been shipped to LSM. The first two C-sections have been joined together (figure \ref{fig:tracker}) and coupled to the first calorimeter wall, forming half of the Demonstrator Module. \textit{In situ} commissioning of this half detector commenced in February 2017. In the following months, the source frame will be shipped to LSM and installed in the middle of the detector, allowing the full Demonstrator Module to be coupled together and commissioned.

\section{Radon mitigation strategy}

\begin{figure*}[t!]
    \centering
    \begin{subfigure}[t]{0.45\textwidth}
        \centering
        \includegraphics[height=1.5in]{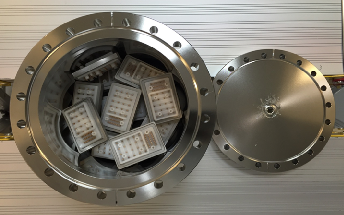}
        \caption{Feedthroughs in emanation chamber}
        \label{fig:emanation}
    \end{subfigure}
    \begin{subfigure}[t]{0.45\textwidth}
        \centering
        \includegraphics[height=1.5in]{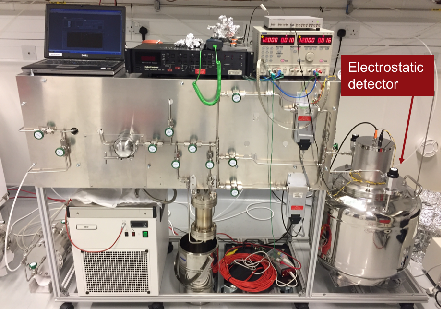}
        \caption{RnCL with electrostatic detector}
        \label{fig:rncl}
    \end{subfigure}
    \caption{Equipment for measuring radon activity levels}
\end{figure*}

One of the largest backgrounds is due to radon, as $\beta$ decays of its daughter isotope $^{214}$Bi can mimic \zeronu{} events in our detector. To protect against this, stringent radiopurity requirements are imposed both on materials used in and around the detector, and on the gas in the tracking chamber.
To measure the activity of detector components, we leave them in an emanation chamber (figure \ref{fig:emanation})  for two weeks, then measure the activity with an electrostatic detector similar to \cite{electrostatic}. To reach SuperNEMO's target sensitivity, the radon activity in the tracker gas must be below 0.15 mBq/m$^3$. As our electrostatic detector is only sensitive to 1~mBq/m$^3$, we flow tracker gas through a Radon Concentration Line (RnCL) \cite{rncl}, capturing any radon in a carbon trap. Measuring the activity of the concentrated gas from the trap lets us calculate tracker activities as low as $10~\mu$Bq/m$^3$. Our measurements indicate that, in order to meet the 0.15 mBq/m$^3$ requirement, we need to flow gas through the Demonstrator Module tracker at the reasonable rate of 2~m$^3$/hour.

\section{Gas system automation}
\begin{figure}[h]
\centering
\includegraphics[height=1.7in]{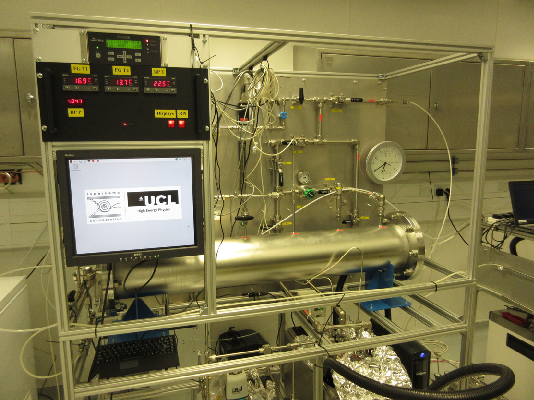}
\caption{SuperNEMO's gas system and electronics}
\label{fig:gas}
\end{figure}
The tracker is filled with a mixture of  95\% helium, 4\% ethanol, and 1\% argon. The gas system (figure \ref{fig:gas})  includes mass flow controllers and a pair of ethanol bubblers used to adjust the gas fractions. A RaspberryPi connected to the slow control system tracks the pressure and temperature in the bubblers and the gas flow rate, and provides alarms and real-time monitoring via a user interface.

\section{Software and analysis}
\begin{figure*}[t!]
    \centering
    \begin{subfigure}[t]{0.45\textwidth}
        \centering
        \includegraphics[height=1.5in]{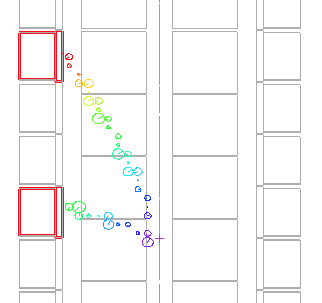}
        \caption{Simulated \zeronu{} event. Circles show drift radii; colours indicate timing.
}
        \label{fig:sim}
    \end{subfigure}
    \begin{subfigure}[t]{0.45\textwidth}
        \centering
        \includegraphics[height=1.5in]{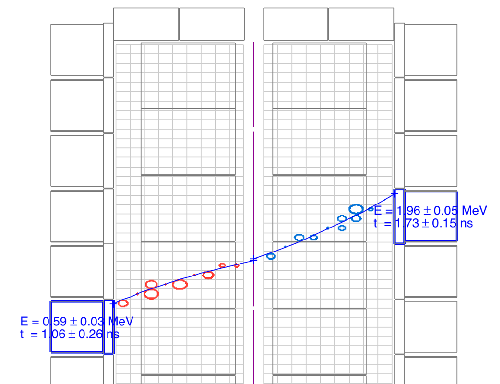}
        \caption{2 reconstructed electrons, showing tracker drift radii and calorimeter hits.
}
        \label{fig:reco}
    \end{subfigure}
    \caption{Simulated \zeronu{} events in the SuperNEMO event viewer}
        \label{fig:flvis}
\end{figure*}
SuperNEMO's simulation and reconstruction software have been used to perform sensitivity studies, confirming initial predictions, and to evaluate the effects of potential sources of background contamination. The event display (figure \ref{fig:flvis}), which visualises and displays information about simulated and reconstructed calorimeter hits, has allowed us to study how signal and background events will present themselves in the detector, enabling us to improve our event selection.  

\section{Conclusion}
SuperNEMO's Demonstrator Module is currently being installed and commissioned at LSM, and will begin taking data in 2017. A stringent radon mitigation strategy gives us ultra low backgrounds, with a  projected Majorana mass sensitivity of 200-400 meV in 2.5 years of running.

\Acknowledgements

\end{document}

%% file: econfmacros.tex
\def\beq{\begin{equation}}
\def\eeq#1{\label{#1}\end{equation}}
\def\eeqn{\end{equation}}

\def\beqa{\begin{eqnarray}}
\def\eeqa#1{\label{#1}\end{eqnarray}}
\def\eeqan{\end{eqnarray}}

\let\bar=\overbar

\def\Dslash{\not{\hbox{\kern-4pt $D$}}}
\def\dslash{\not{\hbox{\kern-2pt $\del$}}}

\def\msb{{\bar{\ssstyle M \kern -1pt S}}}

